\documentclass[12pt,a4paper]{article}
\pdfoutput=1

\usepackage{jheppub}
\usepackage{amsmath,amsfonts,amsthm,amssymb,dsfont}
\usepackage{graphicx,wrapfig,lipsum}
\usepackage[section]{placeins}
\usepackage{slashed}

\usepackage{tikz}
\usetikzlibrary{shapes.geometric,decorations.markings}
\usepackage{subfigure}

\numberwithin{equation}{section}
\newcommand{\Vol}{\text{Vol}}
\newcommand{\Tr}{\text{Tr}}
\newcommand{\diag}{\text{diag}}

\newcommand{\N}{\mathcal{N}}
\newcommand{\D}{\mathcal{D}}
\newcommand{\A}{\mathcal{A}}
\newcommand{\M}{\mathcal{M}}
\newcommand{\ta}{\tilde{a}}

\title{Twisted circle compactification of ${\cal N}=4$ SYM and its Holographic Dual}
\author{S. Prem Kumar and Ricardo Stuardo}
\affiliation{Department of Physics, Swansea University, Swansea, SA2 8PP, U.K.}
\emailAdd{s.p.kumar@swansea.ac.uk, ricardostuardotroncoso@gmail.com}
\abstract{We consider a compactification of 4D $\N=4$ SYM, with $SU(N)$ gauge group, on a circle with anti-periodic boundary conditions for the fermions. We couple the theory to a constant background gauge field along the circle for an abelian subgroup of the $R$-symmetry which allows to preserve four supersymmetries. The 3D effective theory exhibits gapped and ungapped phases, which we  argue are holographically dual, respectively, to a supersymmetric soliton in AdS$_{5}\times S^{5}$, and a particular quotient of AdS$_5\times S^5$. The gapped phase corresponds to an IR 3D ${\cal N}=2$ supersymmetric Yang-Mills-Chern-Simons theory at  level $N$, while the ungapped phase is naturally identified with the root of a Higgs branch in the 3D theory. We discuss the extension of the twisting procedure to maximally SUSY Yang-Mills theories in different dimensions, obtaining the relevant duals for 2D and 6D, and comment on the odd dimensional cases.}

\begin{document}
\maketitle

%---------------------- Main Body -----------------------
\newpage
\setcounter{page}{1}
\setcounter{footnote}{0}

\section{Introduction}

Formulating a QFT with fermions on a manifold with compact directions requires specifying the spin structure. When the compact dimension is a circle, one needs to impose periodic or anti-periodic boundary conditions for the fermions on the circle. In the context of supersymmetric (SUSY) theories, and in particular SUSY Yang-Mills (SYM) theories, choosing anti-periodic boundary conditions breaks SUSY by making all fermion modes massive at the classical level and spoiling Bose-Fermi degeneracy. Any massless scalar superpartners of the fermions then get masses through radiative corrections beginning at one-loop in perturbation theory. At energies much lower than the scale set by the size of the circle, the effective theory is pure Yang-Mills theory in one dimension less. 

In this paper we consider the scenario described above, but with a twist, in theories with extended supersymmetry in various dimensions. In particular, we will discuss circle compactifications with SUSY breaking boundary conditions for fermions, simultaneously switching on a constant abelian background gauge field for a global $R$-symmetry along the compact dimension, in such a way that the low energy effective theory consists of a classically massless SUSY Yang-Mills multiplet preserving at least four supercharges. Quantum effects in the low energy theory can then generate a mass gap, which can be explored quantitatively via a holographic description.

A key motivation for this work is to understand the field theoretic implication of the (holographic) supersymmetric soliton supergravity backgrounds found first in \cite{Anabalon:2021tua} (see \cite{Bobev:2020pjk} for a precursor), and generalisations thereof \cite{Anabalon:2022aig, Nunez:2023nnl, Nunez:2023xgl, Fatemiabhari:2024aua, Anabalon:2024qhf,  Anabalon:2024che}. What makes these backgrounds particularly interesting is that, like holographic duals of SUSY field theories with SUSY-breaking spin structure they contain a cigar-like geometry: shrinking of the circle coordinate along which the boundary field theory is compactified.  Canonical examples \cite{Witten:1998zw} of  such backgrounds are double Wick rotations of non-extremal D$p$-brane geometries, dual to maximally SUSY Yang-Mills in $(p+1)$ dimensions compactified on a circle with anti-periodic (thermal) boundary conditions for the fermions. Crucially however, while the latter backgrounds have vanishing Killing spinors due to the anti-periodicity condition, those in \cite{Anabalon:2021tua, Anabalon:2022aig,Nunez:2023nnl, Nunez:2023xgl, Fatemiabhari:2024aua, Anabalon:2024qhf,  Anabalon:2024che} have non-vanishing Killing spinors satisfying anti-periodic boundary conditions around the circle. A further puzzling feature of the supersymmetric soliton backgrounds of \cite{Anabalon:2021tua} is that unlike other known gravity duals of SUSY field theories with a mass gap \cite{Maldacena:2000yy, Klebanov:2000hb}, there are no additional fields turned on that could correspond to field theory condensates accompanying a nontrivial vacuum structure.

 The $\N=4$ SYM theory on ${\mathbb R}^{1,3}$ provides the most instructive example to focus attention on. Compactifying on a spatial $S^1$ with anti-periodic boundary conditions for fermions, we  first explain how the inclusion of a constant (real) background gauge field, with non-vanishing component along the $S^{1}$, for a diagonal combination of the maximal abelian subgroup of the $SO(6)$ $R$-symmetry, leaves behind a massless gauge supermultiplet  corresponding to a four supercharge theory.

Introducing a constant {\em imaginary} background temporal gauge field for a global abelian symmetry corresponds to turning on a chemical potential for the associated global charge. This has been extensively explored in the context of thermal ${\cal N}=4$ SYM with $R$-symmetry chemical potentials on both ${\mathbb R}^3\times S^1_\beta$ \cite{Yamada:2006rx} and $S^3\times S^1_\beta$ \cite{Hollowood:2008gp} and holographic gravity duals of maximally supersymmetric theories \cite{Yamada:2007gb, Yamada:2008em}. The gravity duals involve $R$-charged black holes \cite{Cvetic:1999ne, Cvetic:1999rb, Chamblin:1999tk, Chamblin:1999hg, Basu:2005pj,Henriksson:2019zph}. The supersymmetric solitons we discuss result from analytically continuing the $R$-charge chemical potentials in those backgrounds to imaginary values and taking the limit of zero energy density in the dual QFT. 

It is important to note that the analytic continuation alluded to above has a drastic effect on the QFT. In particular, ${\cal N}=4$ SYM with a real $R$-charge chemical potential does not have a well defined ground state, which is to say that the grand canonical ensemble is ill defined due to the presence of a flat Coulomb branch moduli space of vacua, leading to brane nucleation instabilities visible  both at weak and strong coupling.  On the other hand, the theory we discuss here with a real background gauge field for an $R$-symmetry is supersymmetric and hence, stable. 

Our main observation is that the twisted compactification of ${\cal N}=4$ SYM can be viewed as a 3D effective ${\cal N}=2$ theory with real mass deformation, which has  gapped and ungapped vacua. Crucially, the gapped vacuum results from the generation of Chern-Simons terms in the 3D effective theory: the spectrum of KK harmonics in the twisted compactification is not vectorlike as explained in \cite{Cassani:2021fyv}, and integrating these out generates a Chern-Simons term with level $N$ in the $SU(N)$ theory. The twisted compactification also allows at the classical level, vacua where new light states appear and lead to Higgs branch moduli spaces. We identify the root of one such branch as the ungapped phase, a quotient of Poincar\'e AdS$_5\times S^5$ with non-shrinking $S^1$, pointed  out in \cite{Anabalon:2021tua}. We provide checks supporting the picture via probe calculations in the dual backgrounds.

This paper is organised as follows: we begin Section \ref{FT} with  a toy example that explains the basic mechanism of the twisted compactification around the circle. We then extend this procedure to $\N=4$ SYM, showing that the compactified theory preserves four Poincar\'e supercharges in 3D. We further elucidate the vacuum structure of the 3D effective theory.
Section \ref{Gravity} is dedicated to the study of the supersymmetric soliton in AdS$_{5}\times S^{5}$ and the accompanying quotiented AdS$_5\times S^5$ geometry. We explain how these describe the compactified field theory, and examine various probes of these backgrounds revealing aspects of the dual field theory. Then, in Section \ref{Extension}, we propose that it is possible to extend the SUSY preserving mechanism to maximally SUSY Yang-Mills in even dimensions by looking at their holographic duals, and conjecture that it is not possible to apply it in odd dimensions. Finally, Section \ref{Conclusion} contains a summary of our results, conclusions and ideas for future research.

\section{
\texorpdfstring{${\cal N}=4$}{N=4} SYM on \texorpdfstring{${\mathbb R}^{2,1}\times S^1$}{R2,1 x S1} with background gauge field}
\label{FT}

 In this section we explain how coupling a SUSY theory to a constant background gauge field for the maximal $U(1)$ subgroups of the $R$-symmetry, allows to preserve some amount of SUSY when imposing SUSY breaking boundary conditions on a spatial circle. We first look at the illustrative example  $\N=1$ SYM in four dimensions and subsequently move to the maximally SUSY theory, which is the main objective of this note.

\subsection{Toy example: \texorpdfstring{${{\cal N}=1}$}{N=1} SYM on \texorpdfstring{${\mathbb R}^{2,1}\times S^1$}{R2,1 x S1}}

Consider the $\N=1$ SYM theory in 4D with some gauge group $G$. The $\N=1$  vector multiplet consists of the gauge field  $A_{\mu}$ and its superpartner gaugino, a Weyl fermion $\lambda$ transforming in the adjoint representation. The Lagrangian of the theory is
    \begin{equation}\label{N=1}
        S_{\N=1} =\frac{1}{g^{2}_{\rm YM}} \int d^{4}x\,   \Tr\left(  -\frac{1}{2} F_{\mu\nu}F^{\mu\nu} 
            - 2i\lambda \sigma^{\mu}D_{\mu}\bar{\lambda} \right),
    \end{equation}
where we define the gauge covariant derivative as $D_{\mu} \equiv \partial_{\mu} - i[A_{\mu},\,\cdot\,]$. We now compactify the theory on a circle of radius $R$ along the $x^3 \equiv \phi$ direction, identifying points under the shift,
\begin{equation}
\phi \sim \phi + 2\pi R,\label{s1phi}
\end{equation}
 with anti-periodic boundary conditions for $\lambda$ and periodic for $A_{\mu}$. This yields mode expansions for the fields around the circle,
    \begin{equation}\label{BCN=1}
        A_{\mu}(x^j,\phi) = \sum_{n\in {\mathbb Z}} e^{i\frac{n}{R}\phi}\left(A^{(n)}_{i}(x^{j}), \,\Theta^{(n)}(x^{j})\right), \qquad 
        \lambda(x^j,\phi) = \sum_{n\in {\mathbb Z}} e^{\frac{i}{R}\left( n+ \frac{1}{2} \right)\phi}\,\lambda^{(n)}(x^{j}),
    \end{equation}
where $\Theta^{(n)}$ are the Kaluza-Klein (KK) modes of $A_\phi$, appearing as scalars in the 3D effective theory, and $i,j= t,x^{1},x^{2}$, the (2+1) Minkowski coordinates. The fermion modes thus acquire effective masses
\begin{equation}
m_{n,\lambda} =  \left|n+\tfrac12\right| R^{-1}\,,\qquad n\in{\mathbb Z}\,.
\end{equation}
Reducing on the circle, the 3D effective theory has a massless gauge field $A_i^{(0)}$ and a classically massless 3D scalar $\Theta^{(0)}$, with  gauge and Yukawa interactions respectively to fermion modes which are all massive, thus breaking SUSY.

The splitting of the bosonic and fermionic perturbative spectrum can potentially be undone by deforming the Lagrangian with a constant background $U(1)$ gauge field ${\cal A}_\mu$ with only non-vanishing component along $\phi$, and which couples only to the fermionic states. This can of course be viewed as a background field for the $U(1)_R$ symmetry of the theory, which is anomalous quantum mechanically. Introducing the background gauge potential is achieved by the replacement
    \begin{equation}
        D_{\mu} \rightarrow \D_{\mu} = D_{\mu} - i q \A_{\mu},\qquad {\A}_\mu = Q \delta^{\phi}_{\mu}\,,
    \end{equation}
    where $q$ is the R-charge of $\lambda$, and the holonomy of the background gauge field around the $S^1$ is,
    \begin{equation}
       \Phi_{\cal R}\equiv \oint_{S^{1}} d\phi \,\A_{\phi} = 2\pi R\, Q\,.
    \end{equation} 
    For the gaugino, $q=+1$, and hence the shifted covariant derivative leads to shifted masses for the fermion modes in the 3D effective theory,
    \begin{equation}
    m_{n,\lambda} \to \left|\frac{n}{R}+\frac{1}{2R} -Q\right| \,.
    \end{equation}
The background field induces a spectral flow, and by setting $Q = 1/2R$ we restore the Bose-Fermi degeneracy of the perturbative spectrum. In this situation the Wilson line around the $S^1$ is $\Phi_{\cal R}=\pi$. The massless perturbative modes correspond to the four supercharge, $\N=2$ SYM multiplet in 3D. Of course, in this case the full quantum theory has an anomaly for the $R$-symmetry which one must worry about. In particular, the effect of the background gauge field can be absorbed in the field redefinition 
    \begin{equation}
        \tilde \lambda = e^{-i q\A_{\phi} \phi} \lambda,
    \end{equation}
where choosing anti-periodic boundary conditions for $\lambda$ leads to periodic boundary conditions for $\tilde\lambda$, but this is then accompanied by a position ($\phi$-)dependent $\theta$-term in the 4D Lagrangian due to the anomaly. In a theory  with a non-anomalous $R$-symmetry, we can expect an operation analogous to the above to lead to a supersymmetric effective 3D description.

\subsection{ \texorpdfstring{$\N=4$}{N=4} SYM on \texorpdfstring{${\mathbb R}^{2,1}\times S^1$}{R2,1 x S1}}

Let us now consider the $SU(N)$ $\N=4$ SYM theory in (3+1)-dimensions. The field content consists of a gauge field $A_{\mu}$, four Weyl fermions $\{\lambda_{a}\}$ ($a=1,...,4$) and six real scalars $\{X^{I}\}$ ($I=1,...,6$), all transforming in the adjoint representation of $SU(N)$. The theory is invariant under the $SO(6)$ $R$-symmetry, under which the scalars transform in the vector representation or ${\bf 6}$ of $SO(6)$, while the fermions transform in the spinor representation i.e. in the fundamental representation ${\bf 4}$ of $SU(4)$.

The $SO(6)\sim SU(4)$ $R$-symmetry has $SO(2)^{3}\sim U(1)^{3}$ as its maximal abelian subgroup. We pick a basis in which the action of each  $SO(2)$ corresponds to a rotation on a specific pair of scalars, which can accordingly be grouped into 3 complex fields,
    \begin{equation}
        Y_{1} = X^{1}+ i\, X^{2}, \quad
        Y_{2} = X^{3}+ i\, X^{4}, \quad
        Y_{3} = X^{5}+ i\, X^{6}.
    \end{equation}
Each $U(1)$ then acts on one of the complex fields $\{Y_{\tilde a}\}$ as a phase rotation. Assembling the fields and their complex conjugates into the  6-vector,
\begin{equation}
\vec{Y}=(Y_{1},Y_{1}^{*},Y_{2},Y_{2}^{*},Y_{3},Y_{3}^{*}),
\end{equation}
the generators in the vector representation of the three $U(1)$ subgroups are given by\footnote{In the real scalar frame, the infinitesimal transformation generated by, for example, ${\cal R}^{\bf 6}_{1}$ is  $\delta X^{1}=-\theta\, X^{2}$, $\delta X^{2}=\theta\, X^{1}$, with $\theta$ the infinitesimal parameter of the transformation.}
    \begin{equation}
        {\cal R}^{\bf 6}_{1} = \diag(1,-1,0,0,0,0),\quad
        {\cal R}^{\bf 6}_{2} = \diag(0,0,1,-1,0,0),\quad
        {\cal R}^{\bf 6}_{3} = \diag(0,0,0,0,1,-1),
    \end{equation}
while the generators in the spinor representation are\footnote{Under ${\cal R}^{\bf 4}_{1}$, $\lambda_{1}\rightarrow e^{i\frac{\theta}{2}}\lambda_{1}$, with $\theta$ the same parameter for the scalar transformation. The factor of 1/2 indicates the spinorial character under the $R$-symmetry.}
    \begin{equation}
        {\cal R}^{\bf 4}_{1} = \frac{1}{2}\diag(1,-1,-1,1),\quad
        {\cal R}^{\bf 4}_{2} = \frac{1}{2}\diag(-1,1,-1,1),\quad
        {\cal R}^{\bf 4}_{3} = \frac{1}{2}\diag(-1,-1,1,1).
    \end{equation}
    
    We take the theory to be  compactified on a circle as before with SUSY breaking boundary conditions, so we have the same mode expansions for the gauge field components $A_i(x^j,\phi)$ and the effective 3D scalar $\Theta(x^j,\phi)$ as in \eqref{BCN=1}. In addition, the four fermions $\{\lambda_a\}$ and six scalars $\{X^I\}$ have Fourier expansions with half-integer and integer modings, respectively:
    \begin{equation}
        X^{I} = \sum_{n} e^{i \frac{n}{R}\phi} X^{I(n)}(x^{i}), \qquad
       \lambda_{a} = \sum_{n} e^{\frac{i}{R}\left(n+\frac{1}{2}\right)\phi} \lambda^{(n)}_{a}(x^{i})\,.
    \end{equation}
 We now couple the theory to a background gauge field ${\cal A}_\mu$ for the diagonal combination of the three $U(1)$ generators discussed above, with non-vanishing component along the compact $\phi$ coordinate, accordingly modifying the gauge-covariant derivatives as,
 \begin{equation}
        D_{\mu} \rightarrow \D_{\mu} = D_{\mu} - i  {\cal R}_{\rm diag}\, \A_{\mu},\qquad\qquad
        \A_\mu = Q\,\delta_\mu^\phi\,.
    \end{equation}
Here ${\cal R}_{\rm diag}$ is the generator of the diagonal $U(1)$ symmetry and the generators are normalised so the complex scalars have unit charge under phase rotations,
    \begin{equation}
        {\cal R}^{\bf 6}_{\rm diag} = \sum^{3}_{n=1} {\cal R}^{\bf 6}_{n} = \diag(1,-1,1,-1,1,-1),\label{rscalar}
    \end{equation}
while for the fermions $\lambda$ (the charge of $\bar{\lambda}$ has the opposite sign)
    \begin{equation}
        {\cal R}^{\bf 4}_{\rm diag} = \sum^{3}_{n=1} {\cal R}^{\bf 4}_{n} = \frac{1}{2}\diag(-1,-1,-1,3).\label{rfermion}
    \end{equation}
The shifted covariant derivatives for each of the fields now imply the following set of effective masses for the charged fields, namely the 3D fermions,
\begin{equation}
    m_{n, \lambda_4}= \left|\frac{n}{R}+\frac{1}{2R} -\frac{3}{2} Q\right|\,,\qquad m_{n, \lambda_{\tilde a}}=\left|\frac{n}{R}+\frac{1}{2R} +\frac{1}{2} Q\right|\,,\quad \tilde a=1,2,3\,.
\end{equation}
and scalars,
\begin{equation}
m_{n,Y_{\tilde a}}=\left|\frac{n}{R} - Q\right|\,,\qquad \tilde a=1,2,3\,,\quad n\in{\mathbb Z}\,.
\end{equation}
We observe that the antiperiodic fermionic field $\lambda_4$ can now have a zero mode precisely when
\begin{equation}\label{RelationQR1}
    Q=\frac1{3R}\,,
\end{equation}
and can be identified as the gaugino for the 3D gauge multiplet. At this value it also follows immediately that the spectra of modes of the three complex scalars $\{Y_{\tilde a}\}$ and their fermionic partners $\{\lambda_{\tilde a}\}$ are {\em degenerate} and {\em massive}.
%From the $\phi$ part of kinetic term of $\bar{\lambda}^{4}$ we have
%    \begin{equation}
%        \left(D_{\phi} + \frac{3i}{2}Q \right)\bar{\lambda}^{4} = \sum_{n}e^{-\frac{i}{R}(n+\frac{1}{2})\phi}\left[ 
%        -i\left(\frac{n}{R}+\frac{1}{2R}-\frac{3}{2}Q\right)\bar{\lambda}^{4(n)} - i \left[A_{\phi},\bar{\lambda}^{4(n)}\right] \right].
%    \end{equation}
%From here, we see that if we choose
 %   \begin{equation}\label{RelationQR1}
%        Q = \frac{1}{3R},
%    \end{equation}
%then, there is no mass term for $\lambda_{4}$. As before, with this choice of $Q$, the background gauge field has a non-trivial holonomy around the $S^{1}$
So the perturbative spectrum is supersymmetric. Unlike the toy example discussed previously, the $R$-symmetry is non-anomalous. The holonomy at the supersymmetric point is 
    \begin{equation}\label{Holonomy}
    \Phi_{\cal R}=    \oint_{S^{1}} d\phi \, {\cal A}_{\phi} = 2\pi R\, Q = \frac{2\pi}{3}.
    \end{equation}
    %For the rest of the fermions we have
%    \begin{equation}
 %       \left(D_{\phi} - \frac{i}{2}Q \right)\bar{\lambda}^{\ta} = \sum_{n}e^{-\frac{i}{R}(n+\frac{1}{2})\phi}\left[ -i\left(\frac{n}{R}+\frac{2}{3R}\right)\bar{\lambda}^{\ta(n)}  - i\left[A_{\phi},\bar{\lambda}^{\ta(n)} \right] \right],
 %   \end{equation}
%where $\ta=1,2,3$. We see that the fermions $\lambda_{\ta}$ get the same mass $m_{\lambda}= (n+2/3)/R$. On the other hand, even though we imposed periodic boundary conditions for the scalars $\vec{Y}$, they also get a mass due to the coupling to the background gauge field
%    \begin{equation}
%        D_{\phi}Y^{\ta} -  i Q Y^{\ta} = \sum_{n} e^{i\frac{n}{R}\phi}\left[ i\left(\frac{n}{R} - \frac{1}{3R}\right)Y^{\ta(n)} - i \left[ A_{\phi}, Y^{\ta(n)}  \right]\right].
%    \end{equation}
%The mass of all the boson modes is then $m^{2}_{Y} = [(n+1/3)/R]^{2}$. Note that the mass term for the $n$-th mode of the bosons is the same as the mass squared of the $(n-1)$-th fermions
%    \begin{equation}\label{Masses}
%        m^{2}_{Y}\bigg{|}_{n} = \frac{1}{R^{2}}\left(n+\frac{1}{3}\right)^{2},\quad
%        m^{2}_{\lambda}\bigg{|}_{n-1} = \frac{1}{R^{2}}\left(n+\frac{1}{3}\right)^{2},
%    \end{equation}
%this means that the coupling the background gauge field $\A_{\mu}$ is to shift the masses of the Kaluza-Klein modes in order to recover a degenerate spectrum between the scalars $T^{\ta}$ and fermions $\lambda_{\ta}$ while also making $\lambda_{4}$ massless. 
The effect of the background gauge field can also be packaged as $\phi$-dependent phases for each of the fields,
    \begin{equation}
        \eta = e^{-i\frac{3}{2}Q\phi}\lambda_{4}, \qquad 
        \eta_{\ta} = e^{\frac{i}{2}Q\phi}\lambda_{\ta}, \qquad
        Z_{\ta} =e^{-iQ\phi}Y_{\ta}.
    \end{equation}
With anti-periodic boundary conditions for all the fermions and periodic for $Y_{\ta}$ the mode expansions for the redefined fields at the supersymmetric point are,
    \begin{eqnarray}\label{BCRedefined}
        &&\eta =  \sum_{n} e^{i \frac{n}{R}\phi} \eta^{(n)}(x^{i}), \\\nonumber
        &&\eta_{\ta} =  \sum_{n} e^{\frac{i}{R}\left( n+\frac{2}{3}\right)\phi} \eta^{(n)}_{\ta}(x^{i}), \\\nonumber
       && Z_{\ta} =\sum_{n} e^{\frac{i}{R}\left( n-\frac{1}{3}\right)\phi} Z_{\ta}^{(n)}(x^{i}).
    \end{eqnarray}   
In this picture we have one periodic fermion, identified as the gaugino, and the rest of the fermions and the scalars have non-trivial winding around the circle, such that we only have periodicity under $\phi \to \phi +6\pi i$ shifts. 
\subsection{4D \texorpdfstring{$\N=1$}{N=1} picture}
It is useful to revisit the argument above in the  4D $\N=1$ language which is natural in this setting. The $\N=4$ theory can be viewed  as an ${\cal N}=1$  vector multiplet coupled to three adjoint chiral multiplets $\{Y_{\tilde a}\}$ where we denote the chiral multiplets with the same symbol as their lowest scalar component, along with the superpotential,
\begin{equation}
    W= \frac{1}{g^2_{\rm YM}} {\rm Tr}\left(Y_1[Y_2,Y_3]\right)\,.
\end{equation}
In this language, the ${\cal N}=1$ theory has a manifest $U(1)_R$ symmetry under which the gaugino $\lambda_4$ has charge $+1$, and the superpotential transforms with charge $+2$. The two other abelian $R$-symmetries of ${\cal N}=4$ appear as  additional independent global $U(1)$ symmetries  which leave $W$ invariant, corresponding to simultaneous and opposite phase rotations of a pair of the three chiral multiplets. If we choose  $U(1)_R$ to act democratically on the three multiplets, they each must have $R$-charge $+\frac23$ in order for $W$ to have $R$-charge 2, and the corresponding fermions have $R$-charge $-\frac13$. Notice that these charge assignments are, up to an overall rescaling of $\frac23$, precisely the assignments in \eqref{rscalar} and \eqref{rfermion}.  

Introducing a background gauge field, $\hat {\cal A}_\mu= \frac{1}{2R} \,\delta_{\mu}^\phi$, for the $R$-symmetry kills the half-integer moding of the gaugino $\lambda_4$, yields the same spectra\footnote{The rescaled charge assignment means that $\hat {\cal A}_\mu$ and ${\cal A}_\mu$  (defined in \eqref{RelationQR1}), are related as ${\cal A}_\phi = \frac23\hat {\cal A}_\phi$ so the holonomy of the background gauge field, and consequently, the spectrum is indeed the same as encountered previously.} and phase redefinitions for the chiral multiplets as discussed above.

 %Obviously, from the $\N=4$ perspective the four fermions are on equal footing, so that it is a choice which fermion to identify with the vector superpartner. However, due the boundary conditions \eqref{BCRedefined}, it is natural to choose the massless fermion $\eta$ to form the vector multiplet $V=(\eta,A_{\mu})$, while the rest of the fermions are paired with the complex scalars to form the three adjoint chiral multiplets $\Phi^{\ta}=(Z^{\ta},\eta_{\ta})$. This is also consistent with the fact that the modes of these field have the same mass spectrum \eqref{Masses}. 

%Continuing with $\N=1$ notation, the action of the $\N=4$ independent supersymmetry generators $Q^{a}$, can be understood as $\N=1$ SUSY transformation in which the vector multiplet is formed by the vector field and one of the four different fermions. For example, the transformation generated by $Q^{1}$ treats the fermion $\eta_{1}$ as the gaugino, whit the rest of the fields are paired as chiral multiplets, while the one generated by $Q^{2}$ pairs $\eta_{2}$ with the vector field, and so on. 

Note that we can reach the same supersymmetric point by compactifying 4D ${\cal N}=4$ SYM on the $S^1$ with SUSY {\em preserving} boundary conditions and subsequently turning on a background gauge field for a global $U(1)$ global symmetry which is {\em not} an $R$-symmetry. Consider the global $U(1)$ which acts on the chiral multiplets as 
\begin{equation}
    Y_{1,2} \to Y_{1,2} \,e^{2i\alpha/3}\,,\qquad Y_3\to Y_3\,e^{-4i\alpha/3}
\end{equation}
leaving the superpotential invariant. Picking a background gauge field $\hat{\cal A}_\mu = \delta_\mu^\phi/2R$ for this $U(1)$ yields the same theory we obtained above with degenerate masses for all matter fields and their harmonics.

Of the four sets of SUSY transformations, the one that is manifest in the ${\cal N}=1$ picture respects the natural pairing that follows from the boundary conditions \eqref{BCRedefined}. Indeed, the SUSY transformations are compatible with the boundary conditions following from the phase redefinitions \eqref{BCRedefined}. This can be seen explicitly from the boundary term obtained from the ${\cal N}=1$ SUSY variation\footnote{Note that for anti-periodic boundary conditions {\em without} the background gauge field, this boundary term would not vanish, signalling the breaking of SUSY.}:
    \begin{eqnarray}  
       \delta_\epsilon S = \int_{\mathbb{R}^{2,1}\times S^{1}} d^{4}x\, &&\partial_{\mu}\Tr\left( D^{\mu}(Z^{\ta})^{\dagger}\epsilon\, \eta_{\ta} - D_{\nu}(Z^{\ta})^{\dagger}\epsilon\, \sigma^{\nu}\bar{\sigma}^{\mu}\eta_{\ta} \right)\\
  &&= 
        \int_{\mathbb{R}^{2,1}} d^{3}x\, \Tr\left( D^{\mu}(Z^{\ta})^{\dagger}\epsilon\, \eta_{\ta} - D_{\nu}(Z^{\ta})^{\dagger}\epsilon\, \sigma^{\nu}\bar{\sigma}^{\mu}\eta_{\ta} \right)\bigg{|}^{\phi=2\pi}_{\phi=0} =0.\nonumber      
\end{eqnarray}

\subsection{Connection to refined index}
On $S^3\times S^1$, the procedure described above mirrors the definition of the refined Witten index  \cite{Cassani:2021fyv} with complexified chemical potentials $(\omega_1,\omega_2)$ which can be thought of  as  complex structure parameters  on  $S^1\times S^3$. The refined index for ${\cal N}=4$ SYM is only periodic under shifts of $\omega_{1,2} \to \omega_{1,2}+6\pi i$  and lives on a three-sheeted cover of the space of complex structures. 
In the notation of \cite{Cassani:2021fyv}, we are  setting $\omega_2\to 0$ and $\omega_1 \to 2\pi i$ which can be viewed as a Cardy-like limit on the second sheet, so that the partition function takes the form
\begin{equation}\label{refined}
Z_{S^3\times S^1} = {\rm Tr} \,e^{-\{Q,Q^\dagger\} + i\pi {\cal R}}\,,
\end{equation}
where fermions are antiperiodic around the $S^1$ and ${\cal R}$ the $R$-charge of the states\footnote{A chiral multiplet with $R$-charge ${\cal R}_b$ contains a fermion with charge ${\cal R}_f={\cal R}_b-1$. This shift cancels the half-integer moding of the fermions, precisely at the special value of the complexified chemical potential in \eqref{refined}.}.

\subsection{Effective 3D \texorpdfstring{$\N=2$}{N=2} SYM with real masses}\label{qft}
Masses obtained by turning on background gauge fields for global $U(1)$'s correspond to real mass deformations of  3D $ {\cal N}=2$ theories \cite{Aharony:1997bx, Cassani:2021fyv}. The lightest KK harmonics in the setup described above fill out the massless ${\cal N}=2$ vector multiplet coupled to three adjoint chiral multiplets with identical real masses
\begin{equation}
    m_1=m_2=m_3=\frac{1}{3R}\,.
\end{equation}
The higher harmonics are however important and their presence has a nontrivial effect on the IR physics as argued below.

The real scalar $\Theta^{(0)}$ in the 3D gauge multiplet is massless and can have a VEV classically. We can always use gauge transformations to diagonalise $\Theta^{(0)}$ so that,
\begin{equation}
\Theta^{(0)}={\rm diag}(a_1, a_2, \ldots a_N)\,,
\end{equation}
where the $\{a_i\}$ are compact scalars, with  identification under shifts\footnote{The normalisation of the shift is fixed by the definition $\Theta^{(0)}=\oint_{S^1} A_\phi d\phi/2\pi R$.} by multiples of $1/R$,
\begin{equation}
    a_i \sim a_i + \frac{1}{R}\,, \qquad \quad 2\pi R\sum_{i=1}^N a_i = 0\,\,{\rm mod}\,\,2\pi \,,
\end{equation}
as such a shift can be absorbed in a relabelling of the Kaluza-Klein harmonics from the compact direction \cite{Davies:1999uw}.
A VEV for $\Theta^{(0)}$ thus breaks $SU(N)$ to $U(1)^{N-1}$, assuming all $Y_{\tilde a}=0$, and there is a classical Coulomb branch moduli space.

%At energy scales $E\ll 1/R$ we obtain 3D $\N=2$ SYM without matter, since the fields $Y_{\ta}$ and $\lambda_{\ta}$ have no massless modes. 
%It is interesting to think further about this theory. The effective 3D gauge coupling for general $N$ is
%\begin{equation}
 %   g^2_{\rm 3D}N =\frac{g^2_{\rm YM}N}{R}\,,
%\end{equation}
%which is dimensionful and defines the dynamical scale of the theory at least when $g^2_{\rm YM}N\ll 1$. 
%These are, of course, the lightest KK harmonics. 

\subsubsection{Gapped vacuum: \texorpdfstring{$SU(N)_N$}{SU(N) N}  \texorpdfstring{$\N=2$}{N=2} SYM}
 Near the origin of the classical Coulomb branch where $\Theta^{(0)}=0$, the $SU(N)$ gauge symmetry is unbroken and  fluctuations of all chiral multiplets are massive. We might then expect that the low energy theory is effectively pure ${\cal N}=2$ SYM in 3D which has a nonperturbatively generated  superpotential \cite{Aharony:1997bx, Davies:1999uw} and potentially runaway vacua in the decoupling limit for massive matter. However, as argued in \cite{Cassani:2021fyv}, the physics is subtler. The twist with the $R$-symmetry yields  a spectrum  for the fermions in the chiral multiplets, $m_{n, \lambda_{\tilde a}} = |n+\frac23|\frac{1}{R}$, $n\in {\mathbb Z}$, which is not vector-like. Integrating out all the fermion KK harmonics produces a Chern-Simons term with a level given formally by  an alternating infinite sum. When regulated, the sum leaves behind a residual Chern-Simons interaction at level $N$ \cite{Cassani:2021fyv,DiPietro:2014bca} for the massless ${\cal N}=2$ vector multiplet. The vacuum is trivially gapped.

  At the perturbative level, the ${\cal N}=2$ vector multiplet (which includes the scalar $\Theta^{(0)}$) acquires a mass $\sim g^2_{\rm YM} N$, lifting the Coulomb branch. As discussed below, we expect $N$ such vacua owing to spontaneous breaking of the ${\mathbb Z}_N$ symmetry for the theory on ${\mathbb R}^{2,1}\times S^1$.

\subsubsection{Higgs branches and massless vacua}
The theory potentially has additional interesting vacua, visible classically, where the gauge group is Higgsed i.e. the unbroken gauge group has rank $r< N-1$.  Such branches appear as a result of the scalars in the  chiral multiplets acquiring non-commuting VEVs. At the roots of such branches, new light states appear for particular values of the 3D scalar $\Theta^{(0)}$. This follows upon examining the form of the 3D scalar potential for $\Theta^{(0)}$ and the chiral multiplets,
\begin{eqnarray}
 V(\Theta^{(0)}, Y_{\tilde a})\,=&& \frac{1}{g_{\rm YM}^2}\sum_{\tilde a =1}^3\sum_n{\rm Tr}\left|[\Theta^{0}, Y_{\tilde a}^{(n)}] + \tfrac1R\left(\tfrac{1}{3}-n\right)\,Y_{\tilde a} ^{(n)}\right|^2 \\\nonumber
&&\qquad\qquad +\frac{1}{g_{\rm YM}^2}\oint_{S^1}{\rm Tr}\left( \left|[Y_1, Y_{2}] \right|^2 +  \left|[Y_1, Y_{3}] \right|^2  + \left|[Y_2, Y_{3}]\right|^2\right)\,,
\end{eqnarray}
where, in the first term, we have made explicit the sum over the KK modes of the chiral multiplet scalars, $Y_{\tilde a}=\sum_n Y^{(n)}_{\tilde a} e^{in\phi/R}$.
\paragraph{Case $N=3$:}
Keeping in mind that eigenvalues of $2\pi R\,\Theta^{(0)}$ must lie in the interval $[-\pi,\pi]$, we point out that something interesting happens for  $SU(3)$ gauge group, when the VEV takes the form,
\begin{eqnarray}\label{higgs0}
2\pi R\,\Theta^{(0)}\, =\,{\rm{diag}}\left(\frac{2\pi}{3}, 0, -\frac{2\pi}{3}\right)\,\equiv \frac{2\pi}{3} J_3\,,
\end{eqnarray}
where $J_3={\rm diag}(1,0,-1)$.
For this choice, there is a precise cancellation of the real mass for some of the fluctuations with $n=0$ or $1$ in $Y_{\tilde a}^{(n)}$ . Each of the three $Y_{\tilde a}$ yield three massless chiral multiplets at this point. In fact, it is fairly easy to show that the scalar potential vanishes with the following VEVs for both the $n=0$ and $n=1$ harmonics of the scalar fields,
\begin{equation}\label{higgs1}
Y^{(0)}_{\tilde a}  =  \begin{pmatrix}
0&0&0\\
Y_{\tilde a,21}&0&0\\
0& Y_{\tilde a, 32}&0
\end{pmatrix}
\qquad  Y^{(1)}_{\tilde a} =  \begin{pmatrix}
0&0&Y_{\tilde a, 13}\\
0&0&0\\
0& 0&0
\end{pmatrix}\,,
\end{equation}
when the VEVs are constrained to satisfy
\begin{equation}
    Y_{2, ij} = \gamma \, Y_{1, ij}\,\qquad  Y_{3, ij} = \tilde\gamma \, Y_{1, ij}\,,\qquad \gamma,\tilde \gamma\in{\mathbb C}\,,
\end{equation}
and the bifundamental labels  $(ij)= (21), (32)$ or $(13)$, matching the non-zero entries in \eqref{higgs2}.
In the $SU(3)$ theory, generic non-zero  VEVs of this type completely Higgs the gauge group. At the origin of this Higgs branch $SU(3)$ is broken to $U(1)\times U(1)$. The matrix of VEVs $Y_{\tilde a} = Y_{\tilde a}^{(0)}+ e^{i\phi} Y_{\tilde a}^{(1)}$ has eigenvalues proportional to the cube roots of unity,
\begin{equation}
   {\rm spec}(Y)= \left(Y_{13}Y_{32}Y_{21}\right)^{1/3} e^{i\phi/3R} \,\left\{1, \,\omega,\, \omega^2\right\}\,,\qquad Y^3 = {\bf 1}_{3\times 3}  \left(Y_{13}Y_{32}Y_{21}\right) e^{i\phi/R}\,.\label{spec}
\end{equation}
Here we have omitted the subscript $\tilde a$ for clarity. Relatedly, 
single trace gauge-invariant operators with non-vanishing classical VEVs are at least cubic in the $\{ Y_{\tilde a} \}$, e.g. ${\rm Tr}\,Y_{\tilde a}^3$, ${\rm Tr}\,Y_1Y_2Y_3$ etc.  In general, traces of monomials of $Y_{\tilde a}$ are non-vanishing only when the number of terms is a multiple of 3.  This indicates that states which are not left invariant  by the simultaneous phase rotations and shift of the $\phi$ coordinate 
\begin{equation}\label{orbifold}
Y_{\tilde a} \to  e^{-2\pi i/3}Y_{\tilde a}\,,\qquad \phi \to \phi +2\pi R\,,
\end{equation}
are projected out.

A natural interpretation of the VEVs, \eqref{higgs1} and \eqref{spec}, as  three dimensional matrices on the Higgs branch, is in terms of the locations of whole D-branes, which can fractionate at the origin of the Higgs branch \cite{Hollowood:2004ek}.

\paragraph{General $N$:} There are two ways of generalising  the $N=3$ example.

First, consider the theory with $N=3k$, $k\in {\mathbb Z}$, where the $Y_{\tilde a,ij}$ are now $k\times k$ blocks and
\begin{equation}\label{higgs2}
  2\pi R\, \Theta^{(0)} = \frac{2\pi}{3}J_3 {\otimes}{\bf 1}_{k\times k}\,.
\end{equation}
At the root of this Higgs branch ($ {\bf H_1} $), the massless states constitute a 3D ${\cal N}=2$,  $S(U(k)\times U(k)\times U(k))$ quiver gauge theory with $3$ sets of massless  bifundamental chiral multiplets $Y_{\tilde a, ij}$, $ (\tilde a =1,2,3)$, for the pair of nodes $(ij)$  in the quiver.

Exploring the Higgs branch with bifundamental VEVs proportional to the identity, $Y_{\tilde a, ij}= y_{\tilde a}\, {\bf  1}_{k\times k}$, only the diagonal $SU(k)$ ${\cal N} =2$ gauge multiplet is unbroken, and we find precisely 3 massless chiral multiplets transforming in the adjoint representation in the low energy theory.  Since KK harmonics of the massless fields on the Higgs branch have integer modings no CS terms will be generated, and we expect this theory to then flow to the 3D ${\cal N}=8$ SCFT in the IR. When $N$ is not a multiple of $3$, i.e. when $N=3k+1$ or $3k+2$, we can still consider classical vacua which consist of $k$ blocks of the kind discussed above, with the remaining entries set to zero. Such vacua would have an additional unbroken $U(1)$ or $U(2)$ factor which we expect to get  gapped due to induced Chern-Simons terms.

There is a second  generalisation  ($ {\bf H_2} $), of the Higgs vacuum seen above for the $SU(3)$ gauge group. Consider again the case with $N=3k$, but with the  real scalar  VEV of the form
\begin{equation}\label{clock}
 \quad 2\pi R\Theta^{(0)}\,=\,\frac{2\pi}{3k}{\rm diag}\left(\frac{3k-1}{2},\frac{3k-3}{2}, \ldots, -\frac{3k-1}{2}\right)\,.
\end{equation}
For $N=3k+1$ or $N=3k+2$ we  add one or two vanishing entries, respectively, to \eqref{clock}. In general we find $3k$ massless perturbative modes for each member of a chiral multiplet $Y_{\tilde a}$. $2k$ of these arise from off-diagonal entries $\sim (Y_{\tilde a}^{(0)})_{k+i, i}$ for $1\leq i\leq 2k$. The remaining  massless fluctuations are those of the $n=1$ harmonic $\sim  (Y_{\tilde a}^{(1)})_{i, i+2k}$ with $1\leq i\leq k$. 

Matrices with these entries non-vanishing, naturally split into $k$ three-dimensional irreducible blocks which each have eigenvalues proportional to the cube roots of unity. These correspond to $k$ whole D-branes exploring the Higgs branch.

\subsection{Polyakov loop} 
The Wilson loop of the $SU(N)$ gauge field around the $S^1$ is a natural gauge-invariant observable in the compactified theory. We will refer to this as the ``Polyakov loop", in a slight abuse of terminology since the Polyakov loop usually denotes the Wilson loop around the thermal circle\footnote{In the Euclidean thermal setting on ${\mathbb R}^3\times S^1$, the Polyakov loop will have a non-zero expectation value which cannot be computed analytically, although it can be computed holographically \cite{Witten:1998zw}. 
 Consistent weak coupling \cite{Aharony:2005bq} and even zero coupling calculations \cite{Aharony:2003sx} can be performed on $S^3\times S^1$ at large-$N$ revealing competing phases with vanishing and non-vanishing expectation values of the Polyakov loop.}:
\begin{equation}
    P =\frac1N{\rm Tr} \,U\,,\qquad U=\,\exp\left(i\oint_{S^1} A_\phi \,d\phi\right)\,,
\end{equation}
with expectation value,
\begin{equation}
    \langle P\rangle = \frac{1}{N}\langle{\rm Tr}\, e^{2\pi i R\Theta^{(0)}}\rangle\,.
\end{equation}
The Polyakov loop is an order parameter for ${\mathbb Z}_N$ center symmetry of the theory on ${\mathbb R}^{1,2}\times S^1$. In the classically trivial vacuum where $\Theta^{(0)}=0$, the Polyakov loop is non-vanishing and spontaneous breaking of centre symmetry then implies $N$ distinct quantum vacua.

When $N=3k$, we can consider vacua where the VEVs consist of $k$ copies of the three dimensional representations, \eqref{higgs2}, and then the Polyakov loop in this Higgs phase is:
\begin{equation}
\langle P\rangle_{\bf H_1} =\,0\,, \quad N=3k, \quad k \in{\mathbb Z}^+\,.
\end{equation}
But since the  eigenvalues of the holonomy matrix $U$ lie clumped at the cube roots of unity, $\frac1N\langle {\rm Tr} U^3\rangle_{\bf H_1}=1$. When $N=3k+1$ or $3k+1$, the vacuum with $k$ three dimensional blocks has additional unbroken gauge symmetries and $\langle P\rangle_{\bf H_1} = {\cal O}(1/N)$. 

More interesting is the classical vacuum $ {\bf H_2} $ with the arrangement \eqref{clock}. For $N=3k$, the Polyakov loop and its multiply wound versions vanish identically in this vacuum, 
\begin{equation}\label{allloops}
    \frac1N\langle {\rm Tr} \,U^m\rangle_{\bf H_2}=0\,, \qquad 1\leq m <3k\,. 
\end{equation}
For $N=3k+1$ and $3k+2$, there will be non-vanishing pieces $\sim 1/N$. In the large-$N$ limit, this vacuum is ${\mathbb Z}_N$ invariant.

\section{Holographic Description}\label{Gravity}

The holographic dual of $\N=4$ SYM compactified on a circle with SUSY breaking spin structure along with a background gauge field was studied in \cite{Anabalon:2021tua}. The key observation of \cite{Anabalon:2021tua} was that two competing non-singular, asymptotically AdS$_5\times S^5$ dual backgrounds exist, one containing  a cigar-like geometry (AdS soliton) with a shrinking circle and another where the circle always remains finite (quotiented Poincar\'e-AdS). Both backgrounds consist of an  $S^5$ fibration over the circle direction. Importantly, the two geometries become degenerate when the asymptotic value of the background gauge field is tuned so that some amount of SUSY is preserved. As we will review below, it is the nontrivial fibration of the $S^5$ that allows to preserve some amount of SUSY.

\subsection{SUSY \texorpdfstring{AdS$_5$}{AdS5} Soliton and Poincaré-AdS quotient}

\paragraph{AdS soliton:} The supersymmetric AdS$_5$ soliton background in type IIB supergravity  is specified by the following form for the metric and five-form flux\footnote{Our parametrisation of the solution differs from that of \cite{Anabalon:2021tua}.}

    \begin{eqnarray}
        &&ds^{2} = \frac{r^{2}}{\ell^{2}}\left( dx^{2}_{2,1} + f(r)d\phi^{2} \right) 
        + \frac{dr^{2}}{\frac{r^{2}}{\ell^{2}}f(r)} 
        + \ell^{2} \sum^{3}_{i=1}[ d\mu^{2}_{i} + \mu^{2}_{i}\left( d\phi_{i} + Q^{3}\ell^{4}\zeta(r)d\phi \right)^{2}]\,,\nonumber\\\label{D3}
        &&F_{5}= G_{5} + \star G_{5}\,,\\\nonumber
        &&G_{5} = -\frac{4}{\ell}\Vol(\M_{5}) + Q^{3}\ell^{4}\sum^{3}_{i=1} d(\mu^{2}_{i})\wedge \left( d\phi_{i} + Q^{3}\ell^{4}\zeta(r)d\phi \right) \wedge dt \wedge dx_{1} \wedge dx_{2} 
    \end{eqnarray}
where $\mu_{i}$ satisfy $\sum_{i}\mu^{2}_{i}=1$, $\Vol(\M_{5}) = \frac{r^{3}}{\ell^{3}}dt\wedge dx_{1} \wedge dx_{2} \wedge d\phi \wedge dr $ and
    \begin{equation}\label{Functions1}
 f(r) = 1 - \frac{Q^{6}\ell^{12}}{r^{6}}, \qquad 
\zeta(r) = \frac{1}{r^{2}} - \frac{1}{Q^{2}\ell^{4}}\,,\qquad \ell =(4\pi g_sN)^{1/4}\,.
    \end{equation}
Here $\ell$ is the AdS radius and that of the $S^5$, and as usual there are $N$ units of five-form flux through the $S^5$. $Q$ is a dimensionful constant with mass dimension $+1$. In this parametrisation, the three $U(1)$ abelian subgroups of the $SO(6)$ symmetry are manifest in the metric via the angular coordinates $\phi_{i}$. The fibration does not introduce extra brane charges. This solution preserves four real supersymmetries.

The function $f(r)$ has one positive root at $r_{0} = Q\ell^{2}$. In the limit $r\rightarrow r_{0}$, the $\phi$ cycle shrinks to zero size. In order for it to shrink smoothly the circumference  $2\pi R$ of the circle is fixed in terms of $Q$, 
    \begin{equation}
        2\pi R = \frac{4\pi}{f'(r_{0})}\frac{\ell^{2}}{r^{2}_{0}}  \quad
        \implies\quad Q=\frac{1}{3R}\,.
    \end{equation}
This is precisely the field theory relation \eqref{RelationQR1} between the magnitude of the background gauge field and the radius of the $S^1$. 

\paragraph{AdS quotient:} For a given $Q$, there is another background with identical asymptopia, which can be viewed as a smooth quotient of AdS$_5 \times S^5$ in the presence of a Wilson line in the $\phi$ direction,
\begin{equation}
ds^2 = ds^2_{{\rm AdS}_5} + \ell^2\sum_{i=1}^3[d\mu_i^2+\mu_i^2\left(d\phi_i-Qd\phi\right)^2]\,.
\end{equation}
The metric and five-form flux have the general form of \eqref{D3} but with $f(r)=1$ and $\zeta(r) = -1/Q^2\ell^4$. The ten dimensional metric is a quotient of AdS$_5\times S^5$ with the identifications,
\begin{equation}
(\phi, \tilde \phi_i) \sim (\phi +2\pi R, \tilde\phi_i -{2\pi}/{3})\,,\qquad \tilde\phi_i \equiv \phi_i - Q\phi\,,
\end{equation}
where the shift of the $\tilde \phi_i$ coordinates is by the background Wilson line $\Phi=2\pi/3$ along the $S^1$. Note that this quotienting precisely mirrors the dual field theory description \eqref{orbifold}  in the Higgs branch \eqref{higgs1}.

Therefore, both backgrounds above in the limit, $r\rightarrow +\infty$, are asymptotically  AdS$_{5}\times S^{5}$, with two key differences. First, the coordinate $\phi$ is compact with period $2\pi R$. The second difference is that in both cases there is a non-vanishing fibration of the $S^{5}$ over the field theory $S^{1}$: 
    \begin{equation}
        ds^{2}(S^{5})\big|_{r\to\infty} = \ell^{2}\sum^{3}_{i=1} [d\mu^{2}_{i} + \mu^{2}_{i} \left( d\phi_{i} - Q d\phi \right)^{2}].
    \end{equation}
This fibration cannot be undone by a gauge transformation of $\phi_{i}$ since the 1-form $\mathcal{A}=Qd\phi$ has a non-trivial holonomy around the $S^{1}$. This is precisely the gauge field used to cancel the SUSY breaking spin structure in the boundary field theory, namely ${\cal N}=4$ SYM on ${\mathbb R}^{1,2}\times S^1$.  Each of the three fibrations corresponds to deforming the boundary theory by three equal background gauge fields $\mathcal{A}=Qd\phi$ for the three $U(1)$ subgroups of the $R$-symmetry. The holonomy $\Phi_{\cal R}$ of this gauge field has to be tuned to $2\pi/3$ to preserve four supercharges. We briefly review how this works below.

\subsection{SUSY Preservation in Supergravity}

How the supersymmetric soliton preserves SUSY was already explained in \cite{Anabalon:2021tua} and \cite{Nunez:2023xgl}. We now schematically review the mechanism for the solution presented here. In type IIB supergravity with only the $F_{5}$ flux turned on, the condition for  vanishing SUSY variations of the gravitino leads to the Killing spinor equation for the SUSY transformation parameter $\epsilon$, 
    \begin{equation}
        \delta\psi_{M} =  D_{M}\epsilon + \frac{i}{16}\slashed{F}_{5} \Gamma_{M}\epsilon = 0\,. 
    \end{equation}
Here $\{x^M\}_{M=0,1\ldots9}$ are spacetime coordinates and $\slashed{F}_{p}=\frac{1}{p!}F_{a_{1}\dots a_{p}}\Gamma ^{a_{1}\dots a_{p}}$ ($a=0,...,9$ are tangent space indices). In pure AdS$_{5}\times S^{5}$ the solution is \cite{Lu:1998nu},
    \begin{equation}
        \epsilon = \begin{pmatrix}
                        1\\
                        0
                    \end{pmatrix} \otimes \epsilon_{\text{AdS}_{5}}\otimes \epsilon_{S^{5}}\,,
    \end{equation}
where $\epsilon_{\text{AdS}_{5}}$ and $\epsilon_{S^{5}}$ are the Killing spinors on AdS$_{5}$ and $S^{5}$ respectively. The AdS$_{5}$ spinor is,
    \begin{equation}\label{KSAdS5}
        \epsilon_{\text{AdS}_{5}} = \sqrt{\frac{r}{\ell}} \epsilon_{+} + \sqrt{\frac{r}{\ell}}\left( \frac{\ell}{r} +  \frac{1}{\ell} x^{\mu}\Gamma_{\mu} \right)\epsilon_{-}\,,
    \end{equation}
where $\epsilon_{\pm}$ are constant spinors satisfying $\Gamma_{\hat{r}}\epsilon_{\pm}=\pm \epsilon_{\pm}$, so that each of them have two independent complex constants. Here, $\epsilon_{+}$ corresponds to field theory Poincaré supercharges, while $\epsilon_{-}$ to super-conformal transformations. On the other hand the Killing spinor on the $S^{5}$ is of the form
    \begin{equation}\label{KSS5}
        \epsilon_{S^{5}} = e^{\frac{i}{2}(\phi_{1} + \phi_{2} + \phi_{3})}\epsilon_{1}(\theta_{1},\theta_{2})
        + \sum^{3}_{i=1} e^{-i\phi_{i}}e^{\frac{i}{2}(\phi_{1} + \phi_{2} + \phi_{3})}\epsilon_{i+1}(\theta_{1},\theta_{2})
    \end{equation}
where $\epsilon_{1},...,\epsilon_{4}$ are the four independent complex spinors and $\theta_{1,2}$ are the polar angles on the $S^5$.

Periodically identifying the spatial coordinate $x^{3}=\phi$ with $\phi \sim \phi + 2\pi R$, with $R$ as in \eqref{s1phi}, and imposing anti-periodic boundary conditions for the Killing spinor along $\phi$ completely breaks SUSY, so there are no non-vanishing solutions.

We can now see how SUSY is preserved by including the constant part of the fibering function $\zeta(r)$ and setting $f(r)=1$ in \eqref{Functions1}, whilst considering the identification $\phi\sim \phi + 2\pi R$.  This is equivalent to coupling the theory on $\mathbb{R}^{2,1}\times S^{1}$ to the background gauge field $\A$ and picking an ungapped IR phase (corresponding to the Poincar\'e-AdS quotient  geometry). The inclusion of $\A$ makes the spinor charged under shifts of the $\phi$ coordinate which appear as phase shifts. One can obtain the resulting spinor by the replacement $\phi_{i}\rightarrow \phi_{i} - Q\phi$ in \eqref{KSS5} 
    \begin{equation}\label{KSS52}
        \epsilon_{S^{5}} = e^{-i\frac{3Q}{2}\phi}e^{\frac{i}{2}(\phi_{1} + \phi_{2} + \phi_{3})}\epsilon_{1}(\theta_{1},\theta_{2})
        + e^{-i\frac{Q}{2}\phi}\sum^{3}_{i=1} e^{-i\phi_{i}}e^{\frac{i}{2}(\phi_{1} + \phi_{2} + \phi_{3})}\epsilon_{i+1}(\theta_{1},\theta_{2})\,,
    \end{equation}
 with $Q=1/3R$. The coefficients of the spinors $\epsilon_{2},...,\epsilon_{4}$ still do not satisfy anti-periodic boundary conditions around the $\phi$, so these must vanish. Similarly the piece of the AdS$_5$ spinor that depends on  $\epsilon_{-}$ in \eqref{KSAdS5}, contains  the term $\phi\,\Gamma_{\phi}$ which does not respect boundary conditions under shifts of $\phi$. The Killing spinor which does satisfy the boundary conditions is of the form
    \begin{equation}
        \epsilon = e^{-i\frac{3Q}{2}\phi}e^{\frac{i}{2}(\phi_{1} + \phi_{2} + \phi_{3})}                        \begin{pmatrix}
                        1\\
                        0
                    \end{pmatrix} \otimes \epsilon_{+} \otimes \epsilon_{1}(\theta_{1},\theta_{2}),
    \end{equation}
wherein we count four real independent spinors, hence the background preserves four supercharges, matching the field theoretic expectation from weak coupling.

In order to obtain the spinor in the supersymmetric soliton background which is  dual to the gapped phase of $\N=4$ SYM with a background gauge field on a circle, we modify the previous expression to have an $r$-dependant spinor in the AdS part $\epsilon_{+} \rightarrow \epsilon_{+}(r)$. This solution also preserves four supercharges. 

The fact that we can we can preserve SUSY by only introducing the constant part of the fibering gauge field is related to the fact that the mechanism by which the AdS-soliton preserves SUSY is very similar to a twisting procedure, as was explained in \cite{Nunez:2023xgl}. There is a twist between the $R$-symmetry with the spin structure on the shrinking circle $\phi$, realised by the non-vanishing fibration of the $S^{5}$ over the $S^{1}$, which allows for the existence of anti-periodic spinors. The $r$-dependent part of the fibration and the function $f(r)$ allow the space to end smoothly at $r=r_{0}$,  and  the one-form $Q^{3}\ell^{4}\zeta(r)d\phi$ to vanish smoothly at the end of the space.

\subsection{Polyakov loop}

The two supersymmetric backgrounds, namely the AdS$_5$ soliton and the (smooth) quotient of Poincar\'e-AdS, are distinguished by two main features. The first difference is that the $\phi$-circle shrinks smoothly in the soliton background, whereas it remains finite sized for all $r$ in the Poincar\'e-AdS quotient. The second, related difference is that the the non-trival fibration function $\zeta(r)$ in the AdS-soliton background leads to a normalisable mode for the 5D effective gauge field \cite{Anabalon:2021tua} which implies a non-zero, constant expectation value for  the $\phi$-component of the $R$-current,
\begin{eqnarray}
&&{\rm AdS \,\,soliton}: \qquad \qquad\qquad\langle J_{\phi}\rangle = \frac{\lambda}{27R^3}\,,\\\nonumber\\\nonumber
&&{\rm Poincar\acute{e}\,\,AdS\,\,quotient}:\quad\,\langle J_{\phi}\rangle = 0 \,,
\end{eqnarray}
where $\lambda = g^2_{\rm YM}N\gg1$ is the 't Hooft coupling of ${\cal N}=4$ SYM. Both backgrounds yield vanishing expectation value for the QFT stress tensor.  This is natural from the point of view of the dual field theory. Unlike ordinary masses,   real masses through the background gauge field deformation naturally  act as a source for the zero mode of the $\phi$ component of the $U(1)_R$ current, so we generically expect
%\begin{equation}
%    \int_{S^1} {\cal L}_{{\cal N}=4} \to \int_{S^1} {\cal L}_{{\cal N}=4} + {\cal A_\phi} \int_{S^1}d\phi J_R^\phi + \mathcal{A}^{2}_{\phi}\int d\phi (Y^{\ta})^{\dagger}Y^{\ta}\,.
%\end{equation}
\begin{eqnarray}
   \frac{1}{2\pi R}\left\langle \int_{S^1}d\phi\, J_R^\phi \right\rangle \neq 0\,.\label{Rcurrent}
\end{eqnarray}

 The shrinking cycle in the AdS-soliton geometry means that we should be able to deduce a non-zero expectation value for the Polyakov-Maldacena loop \cite{Maldacena:1998im}, by wrapping a Euclidean string worldsheet on the cigar geometry \cite{Witten:1998zw}. We can always choose the worldsheet coordinates $(\tau, \sigma)$ such that the string embedding is,
\begin{equation}
        r(\sigma) = \sigma, \quad \phi(\tau) = \tau,
    \end{equation}
with all other coordinates set to constant. The induced metric on the worldsheet is then
    \begin{equation}
        ds^{2}_{\text{ind}} = \frac{dr^{2}}{\frac{r^{2}}{\ell^{2}}f(r)} + \left(\frac{r^{2}}{\ell^{2}}f(r) + \frac{r_0^6}{\ell^2}\zeta(r)^{2} \right)d\phi^{2}\,,
    \end{equation}
and the on-shell worldsheet action (setting $\alpha^\prime=1$), 
    \begin{equation}
        S_{\text{F1}} = \frac{1}{2\pi}\int^{+\infty}_{r_{0}} dr \int^{2\pi R}_{0}  d\phi \sqrt{1 + \frac{r_0^6}{r^{2}f(r)}\zeta(r)^{2}}\,.
    \end{equation}
Using \eqref{Functions1}, $r_0= Q\ell^2$, $Q=1/3R$ and the change of variables $u=r/r_0$,
 %   \begin{equation}
  %      S_{\text{F1}} = R \int^{+\infty}_{r_{0}} dr \sqrt{1 + \frac{r^{6}_{0}}{(r^{6} - r^{6}_{0})}\left(1 - \frac{r^{2}}{r_{0}^{2}}\right)^{2}}.
   % \end{equation}
%From $R=1/3Q = \ell^{2}/3r_{0}$ and using the change of variables $u = r/r_{0}$ we have
we find
    \begin{equation}
        S_{\text{F1}} = \sqrt{\lambda}\, QR\int^{+\infty}_{1} du \sqrt{1 + \frac{\left( u^{2}-1\right)^{2}}{(u^{6} - 1)}}.
    \end{equation}
The integral is regular at  $u=1$, but divergent when $u\rightarrow + \infty$. As is standard in holography, the action of extended probes must be regulated by adding suitable boundary terms \cite{Drukker:1999zq}. In effect, we regulate the integral by cutting it off at $u=\Lambda$, then subtracting the leading divergence, and taking the limit $\Lambda\to\infty$,
    \begin{equation}
        S^{(\text{reg})}_{\text{F1}} = \frac{\sqrt{\lambda}}{3}\lim_{\Lambda\to\infty}\left( \int^{\Lambda}_{1} du \sqrt{1 + \frac{\left( u^{2}-1\right)^{2}}{(u^{6} - 1)}} - \Lambda\right)\approx -0.243\,\sqrt{\lambda}\,,
    \end{equation}
    valid at large 'tHooft coupling $\lambda$. The Polyakov-Maldacena loop is then $P= \exp(-S_{{\rm F1}}^{\rm (reg)})$. The numerical value of the holographic Polyakov loop is not so important. The main observation is that it is non-zero and therefore  the ${\mathbb Z}_N$ symmetry associated to large gauge transformations around the $\phi$ circle is broken\footnote{Note that spontaneous breaking of this ${\mathbb Z}_N$ is not relevant for determining whether the effective 3D theory confines or not.}.  This further implies that there should be $N$ vacua following identical arguments to those presented in the thermal context in \cite{Witten:1998zw}. The domain walls between adjacent vacua correspond to D1-branes along a noncompact spatial direction in the field theory. Such a D1-brane will want to minimise its tension and sit at the tip of the cigar where its tension is $T_{\rm D1}=2N Q^2/\sqrt\lambda$.
    
    It is natural to identify this phase as the IR description of the classical vacuum of the compactified theory with unbroken $SU(N)$ gauge symmetry and a level $N$ Chern-Simons term that trivially gaps the theory. This is consistent with the bulk picture where no additional fields corresponding to gauge theory condensates are turned on, apart from the zero mode of the $R$-current.

For the AdS quotient geometry, the $\phi$-circle is non-shrinking and there is no finite action static string worldsheet configuation  wrapping the circle and extended in the radial AdS direction. The Polyakov loop must therefore vanish. From the gauge theory viewpoint, the candidate for this vacuum is at the root of the Higgs branch ${\bf H_2}$, characterised by the VEVs \eqref{clock}, where the expectation value of the Polyakov loop is always vanishing in the large-$N$ limit \eqref{allloops}.

\subsection{Coulomb branch probes}
We expect that the Coulomb branch of ${\cal N}=4$ SYM is completely lifted in the gauge theory due to real masses for the chiral multiplets, since this deformation only preserves four supercharges. 

\paragraph{D3-brane in the AdS-soliton:} To verify this we introduce a probe D3-brane extended in the field theory directions $x^{\mu}=(t,x^{1},x^{2},\phi)$. The induced metric on the D3-brane is
    \begin{equation}
        ds^{2}_{\text{D3}} = \frac{r^{2}}{\ell^{2}} dx^{2}_{2,1} + \left(\frac{r^{2}}{\ell^{2}}f(r) + \frac{r_0^6}{\ell^2}\zeta(r)^{2}\right)d\phi^{2}, \quad 
        \sqrt{-\det g_{\text{D3}}} = \frac{r^{4}}{\ell^{4}}\sqrt{f(r) + \frac{r_0^6}{r^2}\zeta(r)^{2}}
    \end{equation}
while the pullback of the $C_{4}$ potential is
    \begin{equation}
        C^{*}_{4} = - \left(\frac{r^{4}}{\ell^{4}}-\frac{r_0^4}{\ell^{4}}\right) dt\wedge dx_{1} \wedge dx_{2} \wedge d\phi.
    \end{equation}
Then, the action of the D3-brane probe at fixed radial coordinate has the form,
    \begin{equation}
        S_{\text{D3}} = T_{\text{D}3}\int d^{4}x\, \frac{r^{4}}{\ell^{4}}\left( \sqrt{\left(1-\frac{r_0^2}{r^2}\right)\left(1+\frac{2r_0^2}{r^2}\right)
 %       f(r) + \frac{Q^{6}\ell^{12}\zeta(r)^{2}}{r^{2}}
        } - \left(1-\frac{r_0^4}{r^4}\right) \right) \geq 0\,.
    \end{equation}
    In the AdS$_5$ soliton geometry, the probe D3-branes are therefore attracted to the origin at $r=r_0$, and the Coulomb branch is lifted.
\paragraph{D3-brane in the AdS quotient:} One may naively expect the D3-brane probe potential to vanish in the Poincar\'e AdS quotient geometry. However, this turns out not to be the case since the effect of the twisting is always present and the probe action in this background  is
\begin{equation}
        S_{\text{D3}} = T_{\text{D}3}\int d^{4}x\, \frac{r^{4}}{\ell^{4}}\left( \sqrt{1+\frac{r_0^2}{r^2}} - 1 \right) \geq 0\,.
    \end{equation}
Probe D3-branes of the type considered above have a potential that drags them to the origin.

\subsubsection{D3$^\prime$-brane probe with moduli space}

Now let us consider a D3-brane probe extended in $x^{\mu}=(t,x^{1},x^{2},\phi)$ but such that $\phi_{i} = Q \phi$, that is, the D3-brane wraps the three $U(1)$ subgroups of the $R$-symmetry directions. The metric and four-form potential pulled back onto this D3-brane worldvolume,
    \begin{equation}
        ds^{2}_{{\text{D3}^\prime}} = \frac{r^{2}}{\ell^{2}} dx^{2}_{3,1}, \quad   C^{*}_{4} = 
        -\frac{r^{4}}{\ell^{4}} dt\wedge dx_{1} \wedge dx_{2} \wedge d\phi,
    \end{equation}
exhibit perfect cancellation so that there is a probe moduli space
    \begin{equation}
        S_{\text{D3}^{\prime}} = T_{\text{D3}} \int d^{4}x \sqrt{-g_{\text{D3}^{\prime}}} - T_{\text{D3}}\int C^{*}_{4} 
        = T_{\text{D3}}\int d^{4}x\, \frac{r^{4}}{\ell^{4}}(1-1)=0.
    \end{equation}
This is a particularly interesting configuration, as it effectively undoes the twisting procedure for both the soliton and quotiented Poincar\'e AdS backgrounds. The untwisting is immediately obvious in the latter case, but less trivial for the soliton background. 

We now explain how these additional moduli are consistent with the dual field theory picture presented in section \ref{qft}.  Consider a classical vacuum of the  $SU(N)$ theory where the gauge symmetry is broken to  $SU(N-3)\times U(1)^3$ with a VEV of the form,
    \begin{equation}
        2\pi R\,\Theta^{(0)}={\rm diag}\big(\underbrace{0,0,\ldots 0}_{N-3}\,, \,\tfrac{2\pi}{3},\,0,\, -\tfrac{2\pi}{3}\big)\,.
\end{equation}
    The final three entries follow the arrangement \eqref{higgs0} at the root of a Higgs branch. We can now explore the Higgs branch moduli space emanating from this vacuum by turning on block diagonal VEVs for the remaining scalars $\{Y_{\tilde a}\}$ comprising of a vanishing $(N-3)$-dimensional block, and non-vanishing 3-dimensional block with entries given as in \eqref{higgs1}. This pattern of VEVs breaks the gauge group to $SU(N-3)\times U(1)$. The unbroken $SU(N-3)$ is gapped by the induced Chern-Simons terms, while the leftover $U(1)$ yields a modulus. In the large-$N$ theory, the unbroken $U(1)$ factor and its modulus should be realised naturally via a probe brane with vanishing potential in the dual geometry. This is precisely what we have seen for the AdS-soliton above. 

    The argument can be extended very similarly to the case of the ${\mathbb Z}_N$-symmetric phase \eqref{clock}. This is achieved by considering a block diagonal representation of scalar VEVs consisting of the ${\mathbb Z}_{N-3}$-symmetric block at large-$N$ and a three dimensional probe block of the type described above, dual to a probe D3$^\prime$-brane in the AdS-Poincar\'e quotient background.

\section{Extension to Dp-brane theories (\texorpdfstring{$p=1, 5$}{p=1,5})}\label{Extension}

Given the simple realisation of the twisted circle compactification for the ${\cal N}=4$ theory and its gravity dual, it is natural ask if it is possible to extend this SUSY preserving procedure to different maximally SUSY D$p$-brane theories, particularly those in even dimensions. Consider maximally supersymmetric Yang-Mills theory in $2(l+1)$ dimensions ($l=0,1,2$), formulated on ${\mathbb R}^{1,2l}\times S^1$ with SUSY breaking boundary conditions for the fermions. The $R$-symmetry group of the theory is $SO(8-2l)$ with maximal abelian group $U(1)^{\mathfrak r}$, where ${\mathfrak r}=4-l$, the rank of the $R$-symmetry group. In such cases we expect the background gauge field holonomy to be $\Phi_{\cal R}=2\pi/{\mathfrak r}$. In this paper, we limit ourselves to the holographic descriptions of corresponding field theories, pointing out that the procedure for obtaining the gravity duals to the even dimensional field theories is a straightforward generalisation of the four-dimensional case.

\subsection{Maximally SUSY Yang-Mills in (1+1)}

The supersymmetric soliton background  produced by a large number of  D1-branes has not does appeared in previous works to the best of our knowledge. We present the result for this obtained by  the twisting procedure:
    \begin{equation}\label{D1}
    \begin{aligned}
        ds^{2} &= \frac{r^{3}}{\ell^{3}} \left( -dt^{2}
                       + f(r) d\phi^{2}\right) 
                       + \frac{dr^{2}}{\frac{r^{3}}{\ell^{3}}f(r)}
                     + \frac{\ell^{3}}{r} \left[ \sum^{4}_{i=1}d\mu^{2}_{i} 
                     + \mu^{2}_{i}\left( d\phi_{i} + Q^{2}\ell^{3}\zeta(r) d\phi\right)^{2}\right], \\
        F_{3} &= \frac{6r^{5}}{\ell^{6}}dr\wedge dt\wedge d\phi  
        -Q^{2}\ell^{3} \sum^{4}_{i=1} 
            d(\mu^{2}_{i})\wedge \left( d\phi_{i} + Q^{2}\ell^{3}\zeta(r) d\phi\right) \wedge dt, \\
        \Phi &= \frac{1}{2}\log\left(\frac{\ell^{6}}{r^{6}}\right),
    \end{aligned}
    \end{equation}
where $\Phi$ is the dilaton, and  $\mu_{i}$ satisfy $\sum_{i}\mu^{2}_{i}=1$,
    \begin{equation}
        f(r) =  1 - \frac{Q^{4}\ell^{12}}{r^{8}} ,\quad
        \zeta(r) = \frac{1}{r^{2}} - \frac{1}{Q\ell^{3}},
    \end{equation}
and $\ell = (32\pi^{2}N)^{1/6}$ such that there are $N$ units of $F_{7}=-\star F_{3}$ flux through the $S^{7}$. This solution preserves four supercharges. As in the AdS$_5\times S^5$ case, we have used a parametrisation of the $S^{7}$ that makes manifest the $U(1)^{4}$ maximal abelian subgroup via the coordinates $\phi_{i}$. At $r_{0}=\sqrt{Q \ell^{3}}$ the cycle $\phi$ shrinks to zero size. The period of $\phi$ is
    \begin{equation}
       2\pi R= \frac{\pi}{2Q} \Rightarrow R = \frac{1}{4Q}, 
    \end{equation}
with $R$ the radius of the $S^{1}$ at $r\rightarrow + \infty$. Asymptotically the fibration of the $S^{7}$ on the shrinking cycle does not vanish and cannot be undone by a gauge transformation. It corresponds to the introduction of a background gauge field for the diagonal $U(1)$ of the  $U(1)^4$ Cartan subalgebra of the  $SO(8)$ $R$-symmetry in the boundary field theory. The holonomy of this background field is $\Phi_{\cal R}=\pi/2$. We defer the discussion of the dual $(0+1)$ dimensional effective four-supercharge theory to future work.

\subsection{Maximally SUSY Yang-Mills in (5+1)}

The  supersymmetric soliton geometry dual to the the six dimensional theory on D5-branes compactified on a circle with a twist, was obtained in \cite{Nunez:2023xgl}. Here we present the configuration in a set of variables more suited to the discussion in this paper, and where the maximal abelian subgroup of the $SO(4)$ $R$-symmetry is manifest:
    \begin{equation}\label{D5}
    \begin{aligned}
        ds^{2} &= \frac{r}{\ell}\left( dx^{2}_{4,1} + f(r)d\phi^{2} \right) + \frac{dr^{2}}{\frac{r}{\ell}f(r)} 
            + r\, \ell \left[\sum^{2}_{i=1} d\mu^{2}_{i} + \mu^{2}_{i}\left( d\phi_{i} 
            + Q^{4}\ell^{5}\zeta(r)d\phi \right)\right]\,,\\
        F_{7} &= -\frac{2r}{\ell^{2}}\, dt\wedge dx_{1}\wedge dx_{2}\wedge dx_{3}\wedge dx_{4}\wedge d\phi\wedge dr \\ 
        &\phantom{=} + Q^{4}\ell^{5} \sum^{2}_{i=1} d(\mu_{i})^{2}\wedge  \left( d\phi_{i} + Q^{4}\ell^{5} \zeta(r)d\phi \right)\wedge dt\wedge dx_{1}\wedge dx_{2}\wedge dx_{3}\wedge dx_{4}\,,\\
        \Phi &=  \frac{1}{2}\log\left(\frac{r^{2}}{\ell^{2}}\right)\,.
    \end{aligned}
    \end{equation}
Here $\ell=\sqrt{N}$ such that there are $N$ units of $F_{3}=-\star F_{7}$ through the $S^{3}$ and
    \begin{equation}
        f(r) = 1 - \frac{Q^{8}\ell^{12}}{r^{4}} , \quad \zeta(r) = \frac{1}{r^{2}} - \frac{1}{Q^{4}\ell^{6}}\,.
    \end{equation}
As discussed in \cite{Nunez:2023xgl} the period of the shrinking cycle $\phi$ is independent  of $Q$\,,
    \begin{equation}
        2\pi R = \pi \ell\,.
    \end{equation}
This is traceable to the linear term in $rf(r)$ and the corresponding linear dilaton growth for large $r$, and that the effective theory on the D5-brane is UV-completed by Little String Theory. The value of $Q$ is fixed by the requirement that we have non-vanishing Killing spinors satisfying the antiperiodic boundary conditions around the $\phi$-circle.

The dual field theory is  maximally SUSY Yang-Mills in 6D with $SU(N)$ gauge group, compactified on a circle with a SUSY preserving background gauge field. Curiously, and differently to the other examples, this solution actually preserves eight supercharges. This is because the dual field theory in the IR is five-dimensional, and  minimal SUSY in 5D has eight Poincar\'e supercharges.

%\subsection{Remarks on the odd dimensional cases}
So far, we have only focussed attention on maximally SUSY gauge theories in even dimensions. Holographic duals to these theories, \eqref{D3}, \eqref{D1} and \eqref{D5} realising the twist, can be obtained by truncating the maximal $SO(8-2l)$ gauged SUGRA in $(2l+3)$ dimensions down to the minimal $U(1)$ gauged SUGRA\footnote{This does not apply for the supersymmetric soliton in the D1-brane background, which was found by directly solving the 10D supergravity equations.} (comprising of Einstein-Maxwell theory and possibly a dilaton, originating from the 10D dilaton), where the $U(1)$ gauge field in lower dimensions corresponds to the fibration of the sphere over the shrinking $S^{1}$.

%Compactifying the 5D gauge theory on a circle yields an additional compact scalar in the 4D effective theory. This scalar pairs with the real scalar from the 5D minimal SUSY vector multiplet, and naturally sits in the ${\cal N}=2$ vector multiplet in 4D. The four real scalars sit in the ${\cal N}=2$ adjoint hypermultiplet. A background gauge field along the fifth compact dimension for a linear combination of $U(1)_J$ and $U(1)_R$ can be chosen to leave a massless ${\cal N}=2$ vector multiplet in 4D while rendering the hypermultiplets massive in a way that continues to preserve SUSY. A similar story can play out in the compactification from 3D to 2D (but with four supercharges).

From the gravity dual perspective, it is not entirely clear how the procedure described above would work for odd dimensional field theories. It appears that a simple truncation of the maximal gauged SUGRA in even dimensions to the Einstein-Maxwell-dilaton sector might not be possible, after turning on the $U(1)$ gauge field which implements the twisting/fibering procedure. We leave more detailed investigation of these issues and their dual field theories to future work.
\section{Conclusion and Open Questions}\label{Conclusion}

Motivated by the  gravity backgrounds found in \cite{Anabalon:2021tua}, in this paper we investigated the compactification of $\N=4$ SYM on a circle  with SUSY breaking boundary conditions, including a constant background gauge field for the maximal abelian subgroup of the $R$-symmetry. We understood how this procedure  allows to preserve four supersymmetries, and how the IR physics of the 3D effective ${\cal N}=2$ SYM is in accord with expectation from the holographic dual description in terms of  the supersymmetric AdS soliton and the quotiented AdS$_5\times S^5$ geometries. We also pointed out that it should be possible to extend this mechanism to maximally SUSY theories in other dimensions. For even dimensions, the corresponding dual gravity backgrounds can be readily obtained. For odd dimensional field theories however, we expect the corresponding dual geometries to be more complicated.

There are several interesting directions to explore:
\begin{itemize}
\item{The IR physics of the 3D effective ${\cal N}=2$ SYM  theory is rich. It would be interesting to explore and verify aspects of this through probes in the AdS soliton background which we expect to correspond to a ``mostly gapped" phase with probe Higgs branch flat directions. This is certainly visible in weakly coupled field theory, and we have identified a candidate probe (D3$^\prime$-brane) that appears to explore such flat directions. There are potentially other ``twisted sector" probes  of interest at the roots of  Higgs branches. Another interesting class of observables are Wilson loops in various representions, both around the compact and noncompact directions.}
\item{We have not discussed in detail the weak coupling physics of the 6D and 2D theories upon compactification with twisting. It seems reasonable that the gapped phase in the 6D example must arise from the effective 5D Chern-Simons term, but it is less clear what happens when we compactify the 2D SYM. The holographic duals in each case clearly show that there can be both gapped and ungapped phases.}
    \item In \cite{Nunez:2023xgl} it was suggested that the submanifold $(\phi,r,S^{3})$ of \eqref{D5} preserves an $SU(2)$ structure. It would be interesting to check whether the submanifolds $(\phi,r, S^{5})$ of \eqref{D3} and  $(\phi,r, S^{7})$ of \eqref{D1} also preserve some structure. This could provide more evidence to support the claim that it is not possible to consider a supersymmetric soliton in a D2 or D4-brane background without exciting more fields.
    \item In the toy model of Section \ref{FT}, we argued that for 4D $\N=1$ theories, it is possible to preserve all the SUSY through the twisting procedure. This is questionable when the $R$-symmetry is anomalous. It would be interesting to try to find supersymmetric solitons on backgrounds dual to theories preserving four Poincaré supercharges, and non-anomalous $R$-symmetry, for example AdS$_{5}\times T^{1,1}$, and check whether the resulting background also preserves four SUSY. New families of Type II backgrounds, based on uplifts of the AdS$_{5}$ soliton of \cite{Anabalon:2021tua}, dual to field theories preserving four or more supercharges where studied in \cite{Chatzis:2024kdu,Chatzis:2024top}. 
    \item We have already touched upon a potential difficulty in obtaining supersymmetric solitons in D$p$-brane backgrounds for $p$ even. A more careful analysis of the field theory for these cases may open a route to obtaining such solitons.
\end{itemize}

\section*{Acknowledgements}

We are grateful to Luigi Tizzano for enlightening discussions and drawing our attention to the work of \cite{Cassani:2021fyv}. We would also like to thank Andr\'es Anabal\'on, Mohammad Akhond, Adi Armoni, Nikolay Bobev,  Dimitrios Chatzis, Lewis T. Cole, Jeremy Echeverria Puentes, Ali Fatemiabhari, Timothy J. Hollowood, Carlos Nunez, Marcelo Oyarzo, Neil Talwar, Daniel C. Thompson, Peter Weck and Marcel I. Yáñez Reyes for various discussions and for sharing with us their ideas and knowledge. SPK is supported by STFC Consolidated Grant No. ST/X000648/1. RS acknowledges  support from  STFC grant ST/W507878/1.

{\footnotesize {\bf Open Access Statement} - For the purpose of open access, the authors have applied a Creative Commons Attribution (CC BY) licence to any Author Accepted Manuscript version arising. 

{\bf Data access statement}: no new data were generated for this work.}

\bibliographystyle{JHEP}

\bibliography{ref}

\end{document}